\documentclass{iopart}
\usepackage{graphicx}
\usepackage{dsfont}
\usepackage[hyperindex,
            colorlinks=true,linktocpage=false,
            bookmarks=false,
            breaklinks=true,
            pdftitle={Modelling coloured residual noise in gravitational-wave signal processing},
            pdfauthor={Christian Roever, Renate Meyer, Nelson Christensen},
            pdfkeywords={gravitational wave detectors and experiments, 
                         inverse problems, time series analysis,
                         signal processing, matched filter, 
                         Whittle likelihood, Student-t distribution},
            linkcolor=black,citecolor=black,urlcolor=black]{hyperref}
\usepackage{breakurl}

\providecommand{\realline}{\mathds{R}}
\providecommand{\complexnumb}{\mathds{C}}
\providecommand{\imag}{\mathrm{i}}
\providecommand{\expect}{\mathrm{E}}
\providecommand{\var}{\mathrm{Var}}
\providecommand{\cov}{\mathrm{Cov}}
\providecommand{\invchisq}{\mbox{Inv-}\chi^2}
\providecommand{\indi}[1]{\kappa_{#1}}
\providecommand{\acov}{\gamma}
\providecommand{\re}{\mathrm{Re}}
\providecommand{\im}{\mathrm{Im}}
\providecommand{\powspecD}{S^\ast}
\providecommand{\powspecC}{S}
\providecommand{\intspec}{\mathcal{I}}

\begin{document}

\title[Modelling coloured residual noise\ldots]{Modelling coloured residual noise in~gravitational-wave signal processing}

\author{Christian~R\"over$^1$,
        Renate~Meyer$^2$ and
        Nelson~Christensen$^3$}

\address{$^1$ Max-Planck-Institut f\"{u}r Gravitationsphysik 
         (Albert-Einstein-Institut) 
         and~Leibniz Universit\"{a}t Hannover,
         Hannover, Germany}
\address{$^2$ Department of Statistics, 
         The University of Auckland,
         Auckland, New Zealand}
\address{$^3$ Physics and Astronomy, 
         Carleton College, Northfield, MN, USA}

\begin{abstract}
  We introduce a signal processing model for signals in non-white
  noise, where the exact noise spectrum is a~priori unknown.  The
  model is based on a \textsl{Student's $t$ distribution} and
  constitutes a natural generalization of the widely used normal
  (Gaussian) model. This way, it allows for uncertainty in the noise
  spectrum, or more generally is also able to accommodate outliers
  (heavy-tailed noise) in the data.  Examples are given pertaining to
  data from gravitational wave detectors.
\end{abstract}


\section{Introduction}\label{sec:Intro}
The measurement of gravitational radiation holds great promise for
exciting astro\-nomical observations \cite{Thorne1987,Schutz1999}.
Around the world, efforts are under way to construct and improve
detectors for gravitational waves; among these are the LIGO detectors
in the US \cite{SiggEtAl2008}, GEO \cite{GrotheEtAl2010} and Virgo
\cite{AccadiaEtAl2010} in Europe, and TAMA in Japan
\cite{AraiEtAl2009}, some of which have already started taking
data. Plans for future detectors include advanced LIGO
\cite{HarryEtAl2010}, LCGT in Japan \cite{Kuroda2010}, the Laser
Interferometer Space Antenna (LISA) \cite{DanzmannRuediger2003} and
the Einstein Telescope (ET) \cite{PunturoEtAl2010}. As existing
instruments are becoming increasingly sensitive and future instruments
are approaching completion, sophisticated signal processing methods
are required in order to detect and accurately interpret these often
weak signals within a noisy environment.  For example, signal
detection is usually implemented via matched filtering
\cite{Finn1992,Turin1960,WainsteinZubakov}, while parameter estimation
is commonly done using Bayesian methods
\cite{RoeverMeyerChristensen2007a,SearleSuttonTinto2009}.  Many of the
data analysis procedures employed to date are based on the assumption
of the noise being Gaussian with a known power spectral density
\cite{Finn1992,JaranowskiKrolak2005}.  While these methods have proven
to be computationally efficient and very powerful at discriminating
rare, weak signals within noise, some more flexibility or robustness
is sometimes desired.  One example is the analysis of data to be
expected from LISA, where the measurement noise originates partly from
instrumental as well as astrophysical sources, and the noise spectrum
is only vaguely known a~priori \cite{BarackCutler2004}.

We introduce an approach that was developed in the context of the
latter case, where, along with the signal parameters to be inferred,
the noise's power spectrum needed to be incorporated as an unknown
into the model \cite{Roever2007-thesis,RoeverEtAl2007c}.  The model
developed here turns out to be computationally convenient, as the
additional noise parameters may be analytically integrated out,
leading to a Student-$t$ likelihood expression instead of the original
normal formulation. We expect the same approach to be useful in many
related signal processing contexts, as it allows to specify the prior
information about the power spectrum, i.e., its expected magnitude and
the associated certainty, in a straightforward manner.  In this way,
it should e.g.\ also be able to account for nonstationarities in the
data, in the sense that the power spectrum does not necessarily need
to be assumed to be exactly the same as at an earlier measurement;
other fluctuations, like outliers, could also be accommodated in a
similar manner.  In fact, the resulting model actually falls into a
class of generalizations to the standard Gaussian that have been
proposed for their robustness properties: models accommodating
heavier-tailed noise, or outliers, by either (`directly') implementing
a non-Gaussian noise model or (`indirectly') substituting the
corresponding least-squares procedures by less outlier-sensitive
methods.  Instead of the derivation based on an unknown power
spectrum, the same model may be motivated by assuming the power
spectrum itself to be random, so that the resulting noise
\textsl{mixture distribution} exhibits a greater variability. In the
limiting case of decreasing spectrum variability, the model again
simplifies to the Gaussian.  So the model will also be applicable as
an ad-hoc alternative tunable robust model, with clearly interpretable
``tuning parameters''.

The organization of the paper is as follows.  The time series setup is
introduced in Section~\ref{sec:ModelSetup}.  In the subsequent
sections, the probabilistic modelling is described, including its
time-domain counterpart, the likelihood, prior distributions,
posterior distribution, marginal likelihood and some implications.
Section~\ref{sec:examples} describes the approach using two
illustrative examples of simulated time series with and without a
signal.  The discussion section~\ref{sec:discussion} puts the new
approach into context.  An appendix explicating the Discrete Fourier
Transform conventions used in this paper is attached.

\section{The time series model}\label{sec:Model}
\subsection{The setup}\label{sec:ModelSetup}
Consider a time series $x_1,\ldots,x_N$ of $N$ real-valued
observations sampled at constant time intervals~$\Delta_t$, so that
each observation~$x_i$ corresponds to time~$t_i=i\Delta_t$.  This set
of $N$ observations can equivalently be expressed in terms of
sinusoids of the Fourier frequencies:
\begin{equation}
  x_i = \frac{1}{\sqrt{N\Delta_t}} \sum_{j=0}^{\lfloor N/2 \rfloor} a_j\cos(2\pi f_j t_i) + b_j\sin(2\pi f_j t_i) \label{eqn:SinusoidalNoise} 
\end{equation}
where the variables $a_j$ and $b_j$ each correspond to Fourier
frequencies $f_j=j\Delta_f=\frac{j}{N \Delta_t}$.  The summation
in~(\ref{eqn:SinusoidalNoise}) runs from $j=0$ to $j=\lfloor N/2
\rfloor$, with $\lfloor N/2 \rfloor$ denoting the largest integer less
than or equal to $N/2$.  By definition, $b_0$ is always zero, and
$b_{\lfloor N/2 \rfloor}$ is zero if $N$~is even; going over from
$x_i$'s to $a_j$'s and $b_j$'s again yields the same number ($N$) of
non-zero figures.  The set of ($N$) \textsl{frequency domain}
coefficients $a_j$ and $b_j$ and the \textsl{time domain} observations
$x_i$ are related to each other through a discrete Fourier transform
(and appropriate scaling; 
see the appendix).
The set of trigonometric functions in (\ref{eqn:SinusoidalNoise})
constitutes an orthonormal basis of the sample space, so that there is
a unique one-to-one mapping of the observations in time and in
frequency domain.  Instead of the two amplitudes $a_j$ and $b_j$, the
definition in (\ref{eqn:SinusoidalNoise}) may equivalently be
expressed in terms of a single amplitude and a phase parameter at each
frequency:
\begin{equation}\label{eqn:lambdaPhi}
  x_i \;=\; \frac{1}{\sqrt{N\Delta_t}} \sum_{j=0}^{\lfloor N/2 \rfloor} \lambda_j\, \sin(2\pi f_j t_i + \varphi_j),
\end{equation}
where $\lambda_j = \sqrt{a_j^2+b_j^2}$ and
\begin{equation} 
  \varphi_j=\left\{\begin{array}{ll} \arctan\Bigl(\frac{b_j}{a_j}\Bigr) & \mbox{if } a_j > 0 \\[1ex]
                                     \arctan\Bigl(\frac{b_j}{a_j}\Bigr)\pm\pi &\mbox{if } a_j<0.
                   \end{array}\right. \nonumber
\end{equation}
For each Fourier frequency~$f_j$, let $\indi{j}$ be the number of
Fourier coefficients not being zero by definition, i.e.:
\begin{equation}\label{eqn:indiBetaDefinition}
  \indi{j} = \left\{\begin{array}{ll} 
                       1  & \mbox{if } (j=0) \mbox{ or } (N\mbox{ is even and } j= N/2)\\ 
                       2  & \mbox{otherwise}
                     \end{array}\right.
\end{equation}
so that $\sum_{j=0}^{\lfloor N/2 \rfloor} \indi{j} = N$\@.  Note that
generally one may often simplify to $\indi{j}\!=\! 2\;\forall j$
without introducing a noticeable discrepancy, but in the following we
will stick to the accurate notation of $\indi{}$ depending on the
index~$j$.

For a \textsl{given} time series (either in terms of $x_i$ or $a_j$
and $b_j$), we define the functions of the Fourier frequencies
\begin{eqnarray}\label{eqn:oneSpectrum1}
  p_1(f_j) &=& \frac{a_j^2 + b_j^2}{\indi{j}} 
               \qquad\mbox{and }\\
  p_2(f_j) &=& \frac{a_j^2 + b_j^2}{\indi{j}^2}
               \;=\; \frac{p_1(f_j)}{\indi{j}}
           \;=\; {\textstyle\frac{\Delta_t}{N}}\,|\tilde{x}(f_j)|^2
\end{eqnarray}
for $j=0,\ldots,\lfloor N/2 \rfloor$, where $\tilde{x}$~denotes the
discretely Fourier-transformed time series~$x$, 
as defined in the appendix.
These are the empirical, discrete
analogues of the \textsl{one-sided} and \textsl{two-sided} spectral
power.  The set of $p_1(f_j)$ is also known as the
\textsl{periodogram} of the time series \cite{Jenkins1961}.

Now consider the case where the observations~$x_i$, and consequently
the $a_j$ and $b_j$, correspond to random variables $X_i$, $A_j$
and~$B_j$, respectively.  This may mean that these are realizations of
a random process, where the random variables' probability
distributions describe the \textsl{randomness} in the observations, or
that they are merely unknown, where the probability distributions
describe a \textsl{state of information}, or it may also be a
m\'{e}lange of both \cite{Jaynes}.

The time series has a zero mean if and only if the expectation of all
frequency domain coefficients is zero as well:
\begin{equation}\label{eqn:zeromean}
  \expect[X_i]=0 \quad \forall i 
  \qquad \Leftrightarrow \qquad
  \expect[A_j]=\expect[B_j]=0 \quad \forall j.
\end{equation}
For the probabilistic time series, the spectral power ($p_1(f_j)$ or
$p_2(f_j)$) consequently also is a random variable ($P_1(f_j)$ or
$P_2(f_j)$, respectively).
We denote its expectation value by $\powspecD(f_j)$; if the mean is
zero as in (\ref{eqn:zeromean}), it is given by
\begin{equation} \label{eqn:Spectrum2}
  \powspecD_1(f_j)
   = \textstyle 
   \indi{j} \powspecD_2(f_j) =
   \expect[P_1(f_j)] =
   \expect\Bigl[\frac{A_j^2 + B_j^2}{\indi{j}}\Bigr] 
   \stackrel{\mbox{\small(\ref{eqn:zeromean})}}{=} \textstyle 
 \frac{\var(A_j) + \var(B_j)}{\indi{j}}.
\end{equation}

It is important to note that while the above expectation is closely
related to a time series' power spectral density, it is yet quite
different. Up to here we have considered the discretely Fourier
transformed data and some of its basic properties. The
figure~$\powspecD(f_j)$ refers to data sampled at a particular
resolution and sample size and is defined only at a discrete set of
Fourier frequencies~$f_j$.  One might want to refer to $\powspecD$ as
the \textsl{discretized} power spectrum.  \textsl{If}~in fact the data
are a realization from a stationary random process with (continuous)
power spectral density~$\powspecC(f)$, then the
expectation~$\powspecD(f_j)$ is related to~$\powspecC(f)$ through a
convolution, depending on the sample size~$N$. Only in the limit of an
infinite sample size both of them are equal
\cite{Jenkins1961,Gregory}.  While both continuous and discretized
spectrum are commonly used as an approximation to or in place of one
another, it is crucial to realize that in the following inference will
be done with reference to the case of a finite sample size and the
discretized, convolved spectrum~$\powspecD(f_j)$.

A related useful figure in this context is the \textsl{integrated
spectrum} with respect to some frequency range $[f_1,f_2]$
\cite{JenkinsWatts},
which here we define as
\begin{equation}\label{eqn:intspec}
  \intspec_{[f_1,f_2]} = \Delta_f\sum_{j=j_1}^{j_2} \frac{\indi{j}}{2}\, \powspecD_1(f_j)
\end{equation}
where the summation is done over the corresponding range of Fourier
frequencies ($j_1=\min\{k:f_k>f_1\}$, $j_2=\max\{k:f_k\leq f_2\}$,
where $f_2-f_1\geq\Delta_f$).  The integrated spectrum allows to
investigate or compare (discrete) spectra independent of the
particular sample size~$N$.

\subsection{The normal model}\label{sec:Gaussianity}
Suppose the moments as defined in (\ref{eqn:zeromean}) and
(\ref{eqn:Spectrum2}) are given.  We may set up a corresponding normal
time series model by assuming all the (frequency domain) observables
to be stochastically independent, have zero mean and
\begin{eqnarray}
  \var(A_j) & = & \sigma_j^2,  \\
  \var(B_j) & = & (\indi{j}-1)\,\sigma_j^2  \quad\mbox{for }j=0,\ldots,\lfloor N/2 \rfloor,
\end{eqnarray}
where
\begin{equation}\label{eqn:SigmaAndSpectrum}
  \sigma_j^2
  \; = \; \textstyle \powspecD_1(f_j)
  \; = \; \textstyle \indi{j}\,\powspecD_2(f_j)
  \quad\mbox{for }j=0,\ldots,\lfloor N/2 \rfloor.
\end{equation}
While other choices of variance settings matching the assumptions
would be possible, the assumption of equal variances for $A_j$ and
$B_j$ is the only one that makes the joint density of $(A_j,B_j)$ a
function of overall amplitude $\lambda_j$
alone, independent of the phase~$\varphi_j$, and with that leaves the
model invariant with respect to time shifts.
The same model may also be motivated via the \textsl{maximum entropy
principle}, as the (zero mean, equal variance, independent) normal
distribution maximizes the entropy given the constraints on the
moments given in (\ref{eqn:zeromean}) and (\ref{eqn:Spectrum2}) above
\cite{Bretthorst1999}.
In that way, the normal model constitutes a most convervative model
setup under the given assumptions
\cite{Jaynes,Gregory,Bretthorst1999}.

The joint normal distribution derived here has actually been commonly
used before (see e.g.~\cite{Fisher1929}); the normality assumption
may not only appear as a natural choice, but will also turn out
computationally convenient in the following.  In the context of
Fourier domain data in particular, the normality assumption may also
be motivated via asymptotic arguments, by considering the limit of an
infinite observation time \cite{Brillinger,Champeney,Kawata1966}.
The resulting normal model also is exactly the same as the one
underlying the so-called \textsl{Whittle likelihood}
\cite{Whittle1957,ChoudhouriGhosalRoy2004a}, or the one at the basis
of \textsl{matched filtering}
\cite{Finn1992,Turin1960,WainsteinZubakov}. While intuitively in
other contexts it may often be sufficient to point out the asymptotic
equality $\powspecD(f_j)\approx\powspecC(f_j)$ for
$N\rightarrow\infty$, here it is crucial to appreciate what exactly
the $\sigma_j^2 = \textstyle \powspecD_1(f_j)$ stand for in the case
of a finite sample size~$N$\@.  The difference between the
\textsl{exact} and \textsl{approximate} (``Whittle'') model is
explicated in more detail in~\cite{ChoudhouriGhosalRoy2004a}.

\subsection{The corresponding time-domain model}\label{sec:covariance}
An immediate consequence of the normality assumption for the
frequency-domain coefficients ($A_j$, $B_j$) is that the time-domain
variables ($X_i$), being linear combinations of the frequency-domain
variables, also follow a normal distribution.  The exact joint
distribution of the~$X_i$ is completely determined by their
variance/covariance structure, which may be expressed in terms of the
autocovariance function.  The covariance for any pair of time-domain
observations $X_m$ and $X_n$ (with $m, n \in \{1,\ldots,N\}$, and
corresponding to times $t_m$ and $t_n$) is given by:
\begin{equation}\label{eqn:autocov}
  \cov(X_m,X_n) \nonumber\\
  \;=\;
  \frac{1}{N\Delta_t} 
  \sum_{j=0}^{\lfloor N/2 \rfloor} \left(\powspecD_1(f_j)\,\textstyle\frac{\indi{j}}{2}\,\cos\bigl(2\pi f_j (t_n-t_m)\bigr)\right).
\end{equation}
Note that the autocovariance $\cov(X_m,X_n)=\acov(t_n-t_m)$ depends on
$t_m$ and $t_n$ only via their time lag $t_n-t_m$.  
Remarkably (though not surprisingly), following either the invariance
or the maximum entropy argument to motivate the model structure in
section~\ref{sec:Gaussianity}, the result is a \textsl{strictly
stationary} model for the data \cite{Taqqu1988}.
Note also that $\gamma(0) = \var(X_i) = \intspec_{[0,f_{N/2}]}$.

The autocovariance function allows one to express the distribution of
the~$X_i$ in terms of their ($N\times N$) covariance
matrix~$\Sigma_X$, which is the symmetric, square Toeplitz matrix
\begin{equation}\label{eqn:covmat}
  \Sigma_X\;=\;\left(
  \begin{array}{cccccc}
   \acov(0) & \acov(\Delta_t) & \acov(2\Delta_t) & \cdots  & \acov(\Delta_t) \\
   \acov(\Delta_t) & \acov(0) & \acov(\Delta_t) & \cdots  & \acov(2\Delta_t) \\
   \acov(2\Delta_t) & \acov(\Delta_t) & \acov(0) & \cdots  & \acov(3\Delta_t) \\
   \vdots   & \vdots   & \vdots   & \ddots &  \vdots   \\
   \acov(\Delta_t) & \acov(2\Delta_t) & \acov(3\Delta_t) & \cdots  & \acov(0) \\
  \end{array}
  \right),
\end{equation}
since $\acov$ is periodic such that
$\acov(i\Delta_t)=\acov((N-i)\Delta_t)$.
The periodic formulation in (\ref{eqn:SinusoidalNoise}) makes the
first and last observations $X_1$ and $X_N$ ``neighbours'', just as
$X_1$ and $X_2$ are.  This may seem odd (depending on the context, of
course), but is not an unusual problem, as it arises for any
conventional spectral analysis of discretely sampled data in the form
of spectral leakage \cite{Harris1978}; it may just not always be as
obvious in its consequences.  The above time domain expression sheds
some light on the exact shortcomings of the Whittle likelihood
approximation; you can see for example that it will be a poor
approximation in case of predominantly low-frequency noise and a small
number of observations.  The problem may be tackled via windowing of
the data, or it will also lessen with an increasing sample size;
again, the features of this approximation are discussed in more detail
in~\cite{ChoudhouriGhosalRoy2004a}.

\subsection{Incorporating an unknown spectrum}\label{sec:UnknownSpectrum}
\subsubsection{The likelihood function}
Now suppose the spectrum is unknown and to be inferred from the data
$x_1,\ldots,x_N$.  An unknown spectrum here is equivalent to the
variance parameters $\sigma_0^2,\ldots,\sigma_{\lfloor N/2 \rfloor}^2$
being unknown.  Assuming normality of $A_j$ and $B_j$, the likelihood
function (as a function of the parameters
$\sigma_0^2,\ldots,\sigma_{\lfloor N/2\rfloor}^2$) is:
\begin{eqnarray} 
 &&\; p(x_1,\ldots,x_N\,|\,\sigma_0^2,\ldots,\sigma_{\lfloor N/2\rfloor}^2) \nonumber \\
 \label{eqn:likelihood1}
 & =\; &\;\prod_{j=0}^{\lfloor N/2 \rfloor} \Biggl[ \frac{1}{\sqrt{2\pi}\,\sigma_j} \exp\Bigl(-\frac{a_j^2}{2\sigma_j^2}\Bigr) \times \Bigl(\frac{1}{\sqrt{2\pi}\,\sigma_j}\Bigr)^{\indi{j}-1} \exp\Bigl(-\frac{b_j^2}{2\sigma_j^2}\Bigr)\Biggr]\\
 \label{eqn:likelihood2}
 & =\; &\;\exp\Biggl(-{\textstyle\frac{N}{2}}\log(2\pi)-\sum_{j=0}^{\lfloor N/2 \rfloor}\left[
        \indi{j}\log(\sigma_j)+\frac{a_j^2+b_j^2}{2\sigma_j^2}\right] \Biggr)\\
 \label{eqn:likelihood3}
 & \propto\; &\;\exp\Biggl(-\sum_{j=0}^{\lfloor N/2 \rfloor}
        \frac{\indi{j}}{2}\left[\log\bigl(\powspecD_2(f_j)\bigr)+\frac{\frac{\Delta_t}{N}\,|\tilde{x}_j|^2}{\powspecD_2(f_j)}\right] \Biggr).
\end{eqnarray}
The term $\tilde{x}_j$ here denotes the $j$th element of the
Fourier-transformed data vector, 
as defined in the appendix.
The above likelihood (\ref{eqn:likelihood3}) is exactly equivalent to
the so-called \textsl{Whittle likelihood}, which is an approximate
expression for stationary time series
\cite{Whittle1957,ChoudhouriGhosalRoy2004a}.

If the noise spectrum is a~priori known, the
``$\log\bigl(\powspecD_2(f_j)\bigr)$'' term may also be irrelevant
(see (\ref{eqn:condLikelihood1})), and the model again reduces to the
normal model described e.g.\ in \cite{Finn1992} that is also the
basis for matched filtering \cite{Turin1960,WainsteinZubakov}.  In
that case, the logarithmic likelihood may be conveniently expressed in
terms of an inner product of vectors, allowing for a geometric
interpretation that is expecially useful in the case of linear signal
models \cite{Finn1992,JaranowskiKrolak2005,MGB}.

\subsubsection{The conjugate prior distribution}\label{sec:conjPrior}
For the model defined above, the conjugate prior distribution for each
of the $\sigma_0^2,\ldots,\sigma_{\lfloor N/2 \rfloor}^2$ is the
\textsl{scaled inverse $\chi^2$-distribution} with scale
parameter~$s_j^2$ and degrees-of-freedom parameter~$\nu_j$:
\begin{equation}\label{eqn:sigmaprior}
  \sigma_j^2 \;\sim\; \invchisq(\nu_j, s_j^2)
\end{equation}
with density function
\begin{equation}\label{eqn:invchisqdensity}
  f_{\nu_j,s^2_j}\bigl(\sigma^2_j\bigr)
  \;=\;
  \frac{\Bigl(\frac{\nu_j\,s^2_j}{2}\Bigr)^{\nu_j/2}}{\Gamma(\nu_j/2)}
  \,
  \bigl(\sigma_j^2\bigr)^{-(1+\nu_j/2)}
  \,
  \exp\biggl(-\frac{\nu_j\, s^2_j}{2\,\sigma^2_j}\biggr)
\end{equation}
\cite{BDA}. The degrees-of-freedom here denote the precision in the
prior distribution, while the scale determines its order of magnitude.
For increasing~$\nu_j$ the distribution's variance goes towards zero,
and for $\nu_j\rightarrow 0$ the density converges toward the
non-informative (and improper) distribution with density
$f(\sigma_j^2)=\frac{1}{\sigma_j^2}$, that is uniform on
$\log(\sigma_j^2)$, and which also constitutes the corresponding
\textsl{Jeffreys prior} for this problem \cite{Jeffreys1946}.
If $\sigma_j^2$ follows an $\invchisq(\nu_j,s_j^2)$ distribution, then
its expectation and variance are given
by\begin{equation}\label{eqn:sigmaMoments}\textstyle
\expect[\sigma_j^2]\,=\,\frac{\nu_j}{\nu_j-2}\,s_j^2
\quad\mbox{and}\quad
\var(\sigma_j^2)\,=\,\frac{2\nu_j^2}{(\nu_j-2)^2(\nu_j-4)}\,s_j^4,
\end{equation}
and the mean and variance are finite for $\nu_j>2$ and $\nu_j>4$,
respectively \cite{BDA}.

In addition to the Jeffreys prior (with $\nu=0$) already mentioned
above, other improper prior distributions may be implemented as
special cases of an~$\invchisq(\nu,s^2)$ distribution. A uniform prior
distribution on $\sigma$ corresponds to ($\nu=-1$, $s^2=0$), and a
uniform prior on $\sigma^2$ corresponds to ($\nu=-2$, $s^2=0$).
Generally, a prior with density
$f(\sigma^2)\propto\frac{1}{(\sigma^2)^k}$ (for $k\geq 0$) corresponds
to an~$\invchisq(\nu=2(k-1),s^2=0)$ distribution
\cite{Bolstad2007ch15}. As usual, care must be taken when using such
improper priors, as the resulting posterior may not be a proper
probability distribution \cite{BDA}.

\subsubsection{The posterior distribution}\label{sec:posterior}
Due to the use of a \textsl{conjugate} prior distribution, the
posterior distribution of the $\sigma_j^2$ for given data again is of
the same family. The posterior density is defined by the product of
prior~(\ref{eqn:invchisqdensity}) and
likelihood~(\ref{eqn:likelihood2}):
\begin{eqnarray}\label{eqn:sigmaposterior1}\textstyle
  p\bigl(\sigma_j^2\,\big|\,\{x_1,\ldots,x_N\} \bigr)
  &\propto&
  \bigl(\sigma_j^2\bigr)^{-(1+\nu_j/2)}
  \,
  \exp\biggl(-\frac{\nu_j\, s^2_j}{2\,\sigma^2_j}\biggr)\nonumber\\
  &&
  \times\;
  \bigl(\sigma_j^2\bigr)^{-\indi{j}/2}
  \,
  \exp\biggl(-\frac{a_j^2+b_j^2}{2\,\sigma^2_j}\biggr)\\
  &=&
  \bigl(\sigma_j^2\bigr)^{-(1+\frac{\nu_j+\indi{j}}{2})}
  \,
  \exp\biggl(-\frac{\nu_j\, s^2_j + a_j^2+b_j^2}{2\,\sigma^2_j}\biggr)
\end{eqnarray}
which again can be recognized as a scaled inverse $\chi^2$-density (cp.~(\ref{eqn:invchisqdensity})):
\begin{equation}\label{eqn:sigmaposterior2}\textstyle
  \sigma_j^2\,\Big|\,\{x_1,\ldots,x_N\} \;\sim\; 
  \invchisq\left(\nu_j+\indi{j},\, \frac{\nu_j s_j^2\,+\,a_j^2+b_j^2}{\nu_j\,+\,\indi{j}}\right),
\end{equation}
where all the different parameters corresponding to different
frequencies are mutually independent.  Comparing prior and posterior
parameters ((\ref{eqn:sigmaprior}), (\ref{eqn:sigmaposterior2})), and
the way the prior and likelihood are combined
(\ref{eqn:sigmaposterior1}), one can see that the prior distribution
might be thought of as providing the information equivalent to~$\nu_j$
observations (of coefficients $a_j$ or $b_j$) with average squared
deviation $s_j^2$ \cite{BDA}.  The prior is essentially of the same
functional form as the likelihood (\ref{eqn:sigmaposterior1}), modulo
a $\frac{1}{\sigma^2}$ term, which resembles an overall
(uninformative) \textsl{Jeffreys prior} prefactor
\cite{Jeffreys1946}. The use of the conjugate prior distribution hence
is not only a computationally convenient choice, but may also appear
as a ``natural'' way of expressing prior information in this context.

The use of the conjugate prior distribution leads to a convenient
expression for the posterior distribution of the variance
parameters~$\sigma_j^2$, and with that of the complete discrete
spectrum.  Also, if $\sigma_j^2\sim\invchisq(\nu_j, s_j^2)$, then,
since $s_j^2$ is a scale parameter, it follows that the distribution
of the two-sided spectrum $\powspecD_2(f_j)=\sigma_j^2/\indi{j}$
simply is $\invchisq(\nu_j, s_j^2/\indi{j})$.  There is a
deterministic relationship between given variance
parameters~$\sigma_j^2$ and the implied autocorrelation
function~$\acov(t)$ (see (\ref{eqn:autocov})), and so for
random~$\sigma_j^2$ the distribution of the corresponding~$\acov(t)$
may be numerically explored via Monte Carlo sampling from the
distribution of the~$\sigma_j^2$.  The expectation and variance
of~$\acov(t)$ may also be derived analytically: the expected
autocovariance (\ref{eqn:autocov}), with respect to the distribution
of the~$\sigma_j^2$, is
\begin{equation}\label{eqn:autocovExpect}
  \expect[\acov(t)] \;=\; 
  \frac{1}{N\Delta_t} 
  \sum_{j=0}^{\lfloor N/2 \rfloor} \left(\expect[\sigma_j^2]\, \textstyle\frac{\indi{j}}{2}\,\cos(2\pi f_j t)\right),
\end{equation}
which is finite as long as $\expect[\sigma_j^2]$ is finite for
all~$j$\@.  Similarly, the variance of the autocorrelation is
\begin{equation}\label{eqn:autocovVar}
  \var(\acov(t))\;=\;
  \frac{1}{N^2\Delta_t^2} 
  \sum_{j=0}^{\lfloor N/2 \rfloor} \left(\var(\sigma_j^2)\,\textstyle\frac{\indi{j}^2}{4}\,\cos(2\pi f_j t)^2\right),
\end{equation}
which again is finite as long as $\var(\sigma_j^2)$ is finite for
all~$j$.

\subsubsection{The marginal likelihood}\label{sec:margLikeli}
The use of the conjugate $\invchisq$ prior (with $\nu_j>0$ degrees of
freedom) for the variance parameters~$\sigma_j^2$ allows to integrate
the unknown noise spectrum out of the likelihood expression
(\ref{eqn:likelihood3}).  The \textsl{marginal} likelihood function
then is
\begin{eqnarray}
  & & p(x_1,\ldots,x_N\,|\,\nu_0,\ldots,\nu_{\lfloor N/2\rfloor},s_0^2,\ldots,s_{\lfloor N/2\rfloor}^2) \nonumber 
  \\
  \label{eqn:margLikelihood0}
  &=& \prod_{j=0}^{\lfloor N/2\rfloor}
  \int_0^\infty p(a_j,b_j|\sigma^2_j)\,p(\sigma^2_j|\nu_j,s^2_j)\,\mathrm{d}\sigma^2_j
  \\
  \label{eqn:margLikelihood1}
  &=& \prod_{j=0}^{\lfloor N/2\rfloor}
      \frac{\left(2\,\pi\right)^{-\frac{\indi{j}}{2}}\;\left(\frac{\nu_j s_j^2}{2}\right)^{\frac{\nu_j}{2}}}{\left(\frac{\nu_j s_j^2 + (a_j^2+b_j^2)}{2}\right)^{\frac{\nu_j+\indi{j}}{2}}}
      \,
      \frac{\Gamma\left(\frac{\nu_j+\indi{j}}{2}\right)}{\Gamma\left(\frac{\nu_j}{2}\right)}
  \\
  \label{eqn:margLikelihood5}
  &\propto& \prod_{j=0}^{\lfloor N/2\rfloor} \biggl(1 + \frac{{\indi{j}^2\,\textstyle\frac{\Delta_t}{N}}|\tilde{x}_j|^2}{\nu_j s_j^2}\biggr)^{-\frac{\nu_j + \indi{j}}{2}}
  \\
  \label{eqn:margLikelihood6}
  &=& \exp\Biggl(-\sum_{j=0}^{\lfloor N/2\rfloor} {\textstyle\frac{\nu_j+\indi{j}}{2}}\,\log\biggl(1 + \frac{{\indi{j}^2\,\textstyle\frac{\Delta_t}{N}}|\tilde{x}_j|^2}{\nu_j s_j^2}\biggr)\Biggr),
\end{eqnarray}
which constitutes a product of Student-$t$ densities with
$(\nu_j+\indi{j}-1)$ degrees of freedom \cite{BDA}.  Using the
uninformative, improper Jeffreys prior (with $\nu_j=0$ degrees of
freedom and density $p(\sigma_j^2)\propto\frac{1}{\sigma_j^2}$) yields
the marginal likelihood
\begin{eqnarray} \label{eqn:margLikelihood7}
  p(x_1,\ldots,x_N)
  &\propto& 
  \prod_{j=0}^{\lfloor N/2\rfloor}\bigl(\indi{j}^2\,\textstyle\frac{\Delta_t}{N}\,|\tilde{x}_j|^2\bigr)^{-\frac{\indi{j}}{2}}.
\end{eqnarray}
One may then also get a mixture of the above expressions
((\ref{eqn:margLikelihood5}) and (\ref{eqn:margLikelihood7})) in case
of a prior setting of partly zero and non-zero prior degrees of
freedom.

Note that the corresponding analogue likelihood expression in case of
an a~priori \textsl{known} spectrum (as e.g.\ in
\cite{Whittle1957,Finn1992,ChoudhouriGhosalRoy2004a}, see
also~(\ref{eqn:likelihood3})) was
\begin{equation}
  \label{eqn:condLikelihood1}
  p(x_1,\ldots,x_N|\sigma_0^2,\ldots,\sigma_{\lfloor N/2\rfloor}^2)
  \;\propto\;\exp\Biggl(-\sum_{j=0}^{\lfloor N/2 \rfloor}
        \frac{\indi{j}^2\,\frac{\Delta_t}{N}\,|\tilde{x}_j|^2}{2\,\sigma_j^2} \Biggr),
\end{equation}
so that the (normal) sum-of-squares expression
(\ref{eqn:condLikelihood1}) in case of a known spectrum generalizes to
the Student-$t$ expression (\ref{eqn:margLikelihood6}) once one takes
uncertainty about the spectrum's scale into consideration.  The normal
model in turn is the limiting case for increasing degrees of freedom.

\subsection{Robust inference via the Student-$t$ model}\label{sec:StudentT}
The Student-$t$ marginal likelihood
expression~(\ref{eqn:margLikelihood5}) suggests that considering noise
as normal but with unknown spectrum is technically equivalent to
viewing the noise itself as being $t$-distributed.  In that way, the
Student-$t$ model may also be useful as an alternative for robust
modelling, as the wider family of $t$-distributions includes the
normal and Cauchy distributions as special or limiting cases.  In
other contexts the $t$-distribution is commonly used as a
generalization of the normal model in order to accommodate
heavier-tailed errors (see e.g.\
\cite{LangeEtAl1989,Geweke1993,Divgi1990,McDonaldNewey1988}).  In contrast to
the above derivation of the $t$-distributed noise based on
prior/posterior distributions, the noise may also be considered as a
\textsl{scale mixture} of normal distributions; that is, normal noise
with a randomly varying scale (variance) \cite{Geweke1993}.  Both
relations may actually be used to motivate the use of the Student-$t$
distribution for noise exhibiting certain nonstationarities (e.g., a
noise spectrum that is slightly fluctuating over time), be it to model
the ``variability'' or the ``uncertainty'' in the spectrum, both of
which are mathematically the same here (except that uncertainty is
reduced through learning, while the random variation is not).

The use of heavier-tailed noise models already has repeatedly been
advocated in the gravitational wave detection context, in order to
gain robustness against outliers and similar deviations from the
commonly assumed normality.  Creighton~\cite{Creighton1999} for
example suggests the use of a mixture of Gaussian noise and a
(uniform) burst component in order to account for wide-band noise
burst events.  Allen et al.\ \cite{AllenEtAl2002} also illustrate the
use of a two-component Gaussian mixture in order to reduce the
influence of outliers in the data, but more generally they advocate an
approach also known as \textsl{M-estimation}, namely the explicit
downweighting or ignorance of outlier observations falling far into
the tail of the noise distribution \cite{Hampel,Huber}.  Application
of the above Student-$t$ model may also be considered as a special
case of robust M-estimation, based on a clearly interpretable noise
model and associated parameters \cite{Divgi1990,McDonaldNewey1988}.

In case of an a~priori known power spectrum, maximization of the
normal likelihood (\ref{eqn:condLikelihood1}) is the basis for the
\textsl{matched filtering} approach commonly used in signal detection
problems, when looking for signals of parameterized shape in noise
\cite{Turin1960,WainsteinZubakov,JaranowskiKrolak2005}. In Gaussian
noise, the likelihood maximization then is equivalent to a
least-squares approach.  A filter based on the Student-$t$ likelihood
(\ref{eqn:margLikelihood5}) may therefore be useful for cases of an
unknown noise spectrum or non-Gaussian noise.  

Note that since the likelihood (\ref{eqn:margLikelihood5}) implies
\textsl{independent} (as opposed to merely uncorrelated) errors,
likelihood maximization here is different from least-squares
estimation
\cite{KelejianPrucha1985,BreuschRobertsonWelch2001,Geweke1993}, and
the likelihood might in fact exhibit multiple modes
\cite{MakelainenEtAl1981}.  The (independent) Student-$t$ model not
only leads to a less drastic fall-off of the likelihood for extreme
values, and hence reduced leverage of outliers, but it also implies
non-spherical density contours for the joint distribution of the
noise, effectively allowing for a fraction of scrambled
\textsl{individual} noise residuals.  A similar effect was pointed out
by Creighton~\cite{Creighton1999} when implementing a robust
Gaussian/uniform mixture noise model: the approach would allow for
excess noise in individual interferometers, so that a noise burst
would essentially be automatically ``vetoed'' if it is only measured
in one of several interferometers. Similarly, the non-spherical
density contours of the Student-$t$ model make it robust against odd
data values at individual frequencies.

\subsection{Defining the prior distribution's parameters}\label{sec:priorparameters}
Depending on the particular application and context in mind, there may
be different ways to sensibly specifying prior parameters.  
Firstly, there are the supposedly uninformative priors; the
\textsl{Jeffreys prior} \cite{Jeffreys1946} with $\nu_j=0$ degrees of
freedom, and priors that are uniform on~$\sigma_j$ or on~$\sigma_j^2$
were already mentioned above (see Sec.~\ref{sec:conjPrior}).
Care needs to be taken here though, as these
priors (with~$\nu_j\leq 0$) are improper distributions. These may lead
to improper marginal likelihoods, as in the case of
(\ref{eqn:margLikelihood7}), which does not correspond to a
normalizable probability distribution for the noise. The resulting
posterior distribution of signal or noise parameters then \textsl{may}
or \textsl{may not} be a proper probability distribution \cite{BDA}.

If the prior information on the spectral parameters is in the form of
\textsl{measurements} (or \textsl{samples}) of the spectrum (of same
size and resolution), then this may be expressed in terms of an
equivalent sample size and corresponding prior degrees-of-freedom, as
suggested in Sec.~\ref{sec:posterior}\@. Choosing e.g.\ the $\nu_j$ to
be twice the number of measurements of the spectrum taken (i.e., equal
to the number of observed Fourier coefficients) would then be
equivalent to \textsl{initially} assuming an uninformative Jeffreys
prior and then using the posterior based on the measurements as the
prior for the actual (signal) analysis. This will of course only make
sense if one assumes the spectrum unknown but stationary.

When using the Student-$t$ model as a robust model accommodating for
heavy-tailed noise, i.e., when the noise itself is assumed to actually
follow a $t$-dis\-tri\-bu\-tion that one can sample from, then one can
use sample estimates for the Student-$t$ parameters. Moment estimators
for $s^2_j$ and $\nu_j$ (based on sample variance and kurtosis) are
given e.g.\ in \cite{SutradharAli1986,Singh1988}.

When specifying prior parameters that are supposed to reflect
information and/or variability, it may be helpful to consider the
implied moments or quantiles, for individual frequency bins or
frequency bands.  The expressions for the prior's moments in
(\ref{eqn:sigmaMoments}) may be inverted to
\begin{equation}\label{eqn:sigmaMomentsInverted}
  \nu_j\;=\;4+2\,\frac{\expect[\sigma_j^2]^2}{\var(\sigma_j^2)}
  \quad\mbox{and}\quad
  s_j^2 \;=\; \frac{\nu_j-2}{\nu_j} \, \expect[\sigma_j^2],
\end{equation}
which allows one to specify the scale~$s_j^2$ and
degrees-of-freedom~$\nu_j$ based on pre-defined prior expectation and
variance of $\sigma_j^2$, respectively.  Note that the
degrees-of-freedom~$\nu_j$ then are simply a function of the prior
variation coefficient~$\sqrt{\var(\sigma_j^2)} \Big/
\expect[\sigma_j^2]$.  A specification of $s_j^2$ independent of~$j$
means a~priori white noise, and specifying individual $\nu_j$ for
different~$j$ indicates varying prior certainty across the spectrum.
A sensible definition of the prior certainties for the individual
spectrum parameters may be complicated by the fact that the exact
meaning of this discrete set of parameters depends on the sample
size~$N$.  In that case it may be helpful to instead consider the
\textsl{integrated spectrum} (\ref{eqn:intspec}) and its a~priori
properties.  The (prior) moments of the integrated spectrum are given
by
\begin{equation}
  \expect[\intspec_{[f_1,f_2]}] 
    \;=\; \Delta_f\sum_{j=j_1}^{j_2} {\textstyle\frac{\indi{j}}{2}} \, \expect[\sigma_j^2]
    \;=\; \Delta_f\sum_{j=j_1}^{j_2} \frac{\indi{j}}{2} \, \frac{\nu_j}{\nu_j-2}s_j^2
\end{equation}
(if all $\nu_j>2$), and
\begin{eqnarray}
  \var(\intspec_{[f_1,f_2]}) 
  &=& \Delta_f^2\sum_{j=j_1}^{j_2} \frac{\indi{j}^2}{4} \, \frac{2\nu_j^2}{(\nu_j-2)^2(\nu_j-4)}s_j^4
\end{eqnarray}
(if all $\nu_j>4$).  The (prior) variation coefficient for the power
within any frequency range then is:
\begin{eqnarray}
  \frac{\sqrt{\var(\intspec_{[f_1,f_2]})}}{\expect[\intspec_{[f_1,f_2]}]} &=&
\frac{\sqrt{\sum_{j=j_1}^{j_2} \frac{\indi{j}^2}{4} \, \frac{2\nu_j^2}{(\nu_j-2)^2(\nu_j-4)}s_j^4}}{\sum_{j=j_1}^{j_2} \frac{\indi{j}}{2}\, \frac{\nu_j}{\nu_j-2}s_j^2}
\end{eqnarray}
which simplifies in case all d.f.\ parameters are taken to be equal
($\nu_j\equiv\nu$):
\begin{eqnarray}
  \frac{\sqrt{\var(\intspec_{[f_1,f_2]})}}{\expect[\intspec_{[f_1,f_2]}]} 
  &=& 
  \sqrt{\frac{2}{\nu-4}}
  \frac{\sqrt{\sum_{j=j_1}^{j_2} \frac{\indi{j}^2}{4}\, s_j^4}}{\sum_{j=j_1}^{j_2} \frac{\indi{j}}{2}\, s_j^2}
\end{eqnarray}
and simplifies further in case all scale parameters are taken to be
equal ($s_j^2 \equiv s^2$):
\begin{eqnarray}
  \frac{\sqrt{\var(\intspec_{[f_1,f_2]})}}{\expect[\intspec_{[f_1,f_2]}]} &=&
  \sqrt{\frac{2}{\nu-4}}
  \frac{\sqrt{\sum_{j=j_1}^{j_2} \frac{\indi{j}^2}{4}}}{\sum_{j=j_1}^{j_2} \frac{\indi{j}}{2}}.
\end{eqnarray}
In the general case of ($j_1>0$, $j_2<\lfloor N/2 \rfloor$) this is:
\begin{eqnarray}\label{eqn:priorDfFreqRange}
  \frac{\sqrt{\var(\intspec_{[f_1,f_2]})}}{\expect[\intspec_{[f_1,f_2]}]} &=&
  \sqrt{\frac{2}{\nu-4}}
  \frac{1}{\sqrt{j_2-j_1+1}}
\end{eqnarray}
and similarly, in case ($j_1=0$, and $j_2=N/2$, $N$~even) it is:
\begin{eqnarray}\label{eqn:priorDfAllFreq}
  \frac{\sqrt{\var(\intspec_{[f_0,f_{N/2}]})}}{\expect[\intspec_{[f_0,f_{N/2}]}]} &=&
  \sqrt{\frac{2}{\nu-4}}
  \frac{\sqrt{\frac{N-1}{2}}}{\frac{N}{2}}.
\end{eqnarray}
This may be useful if one wants to specify piecewise constant prior
settings with given constraints on the overall power \textsl{per
frequency range}; this would then lead to the d.f.\ settings
\begin{eqnarray}
  & \nu\;=\; \textstyle 4+\frac{2}{j_2-j_1+1}\,\Bigl(\textstyle\frac{\expect[\intspec_{[f_1,f_2]}]}{\sqrt{\var(\intspec_{[f_1,f_2]})}}\Bigr)^2\\
\mbox{or}\quad & \nu\;=\; \textstyle 4+2\,\frac{N-1}{N^2}\,\Bigl(\frac{\expect[\intspec_{[f_0,f_{N/2}]}]}{\sqrt{\var(\intspec_{[f_0,f_{N/2}]})}}\Bigr)^2
\end{eqnarray}
respectively, for the above two cases ((\ref{eqn:priorDfFreqRange}),
(\ref{eqn:priorDfAllFreq})).

For example, if one wants the spectrum to be a~priori white with some
scale~$s^2$, such that the marginal prior mean and variation
coefficient of the integrated power
$\intspec_{[f_0,f_{N/2}]}=\var(X_i)=\acov(0)$ are given by
\begin{equation}
  \expect\bigl[\intspec_{[f_0,f_{N/2}]}\bigr] \;=\; \varsigma^2
  \quad\mbox{and}\quad
  \frac{\sqrt{\var\bigl(\intspec_{[f_0,f_{N/2}]}\bigr)}}{\expect\bigl[\intspec_{[f_0,f_{N/2}]}\bigr]} \;=\; c,
\end{equation}
then a setting of 
\begin{equation}
  \nu_j \;=\; 4+{\textstyle\frac{N-1}{N^2}}\,{\textstyle\frac{2}{c^2}}
  \quad\mbox{and}\quad
  s^2_j \;=\; 2\,\Delta_t\,{\textstyle\frac{\nu-2}{\nu}} \, \varsigma^2
\end{equation}
independent of $j$ will yield an a~priori white spectrum with constant
expectation and variation in the overall
variance~$\intspec_{[f_0,f_{N/2}]}$ for any sample size~$N$.
Piecewise constant settings across different frequency bands may be
implemented analogously.  In case all $\nu_j>4$, the integrated power,
being a sum of random variables with finite mean and variance, will be
asymptotically normally distributed.

Instead of using moments, the prior may also be specified in terms of
quantiles, for example by first deciding on the degrees-of-freedom
settings and then aiming the prior median at a certain value.  The
quantiles of an $\invchisq(\nu,s^2)$ distribution may be derived based
on the quantiles of a $\chi^2$-distribution; the $p$-quantile is then
given by
\begin{equation}
  \nu\,s^2 \,\big/ \, \chi^2_{\nu;1\!-\!p}
\end{equation}
where $\chi^2_{\nu;1\!-\!p}$ is the $(1-p)$-quantile of a
$\chi^2$-distribution with $\nu$ degrees of freedom.

Finally, in the M-estimation context, the $t$-distribution's d.f.\
parameter may also be set based on the shape of the corresponding
\textsl{influence function} \cite{Hampel,Huber}.

\section{Examples}\label{sec:examples}
\subsection{Noise only}\label{sec:example1}
Consider a time series~$n(t)$ of $N\!=\!100$ data points sampled at
times~$t_i\!=\!\frac{i}{100}$.
\begin{figure}
  \begin{center}
    \includegraphics[width=0.95\columnwidth]{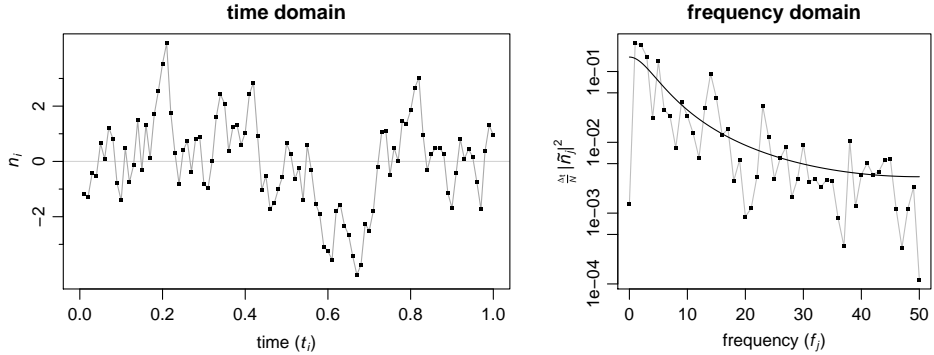}
    \caption{The example data $n_i$ in the time domain, and its
             (empirical) power $\frac{\Delta_t}{N}|\tilde{n}_j|^2$ in
             the Fourier domain. The solid line in the right panel
             shows the theoretical (2-sided) power spectral density in
             comparison.}
    \label{fig:example01}
  \end{center}
\end{figure}
As a simple example of non-white noise, the data are generated from an
autoregressive process
\begin{equation}\label{eqn:AR1noise}
  n(t_i) \;=\; \textstyle \frac{3}{4}n(t_{i-1}) + x(t_i),
\end{equation}
where the innovations $x(t_i)$ are drawn independently from a uniform
distribution across the interval $[-\sqrt{3}, \sqrt{3}]$, so that they
have zero mean and unit variance.  The overall variance of the process
defined in this way
is~$\var\bigl(n(t_i)\bigr)=2.29$; due to the positive
correlation of subsequent samples it has a higher power at low
frequencies and less at high frequencies.  Figure~\ref{fig:example01}
shows a noise sample generated using the above prescription, together
with its theoretical power spectral density.

We will now apply the noise model introduced above and derive the
posterior distribution of the noise parameters.  Assuming one has a
rough idea of the noise variance, we set the prior scale
parameters~$s^2_j$ so that the noise is a~priori white with a prior
expectation of $\expect\bigl[\var\bigl(n(t_i)\bigr)\bigr]=2.50$.
We use $\nu_j=3$ degrees of freedom, so that the noise parameters'
(and with that, the overall power's) prior expectations are finite,
while the variances are not.  Prior scale and prior expectation then
are $s^2_j=0.0166$, and $\expect[\sigma_j^2]=0.05$.
\begin{figure}
  \begin{center}
    \includegraphics[width=0.95\columnwidth]{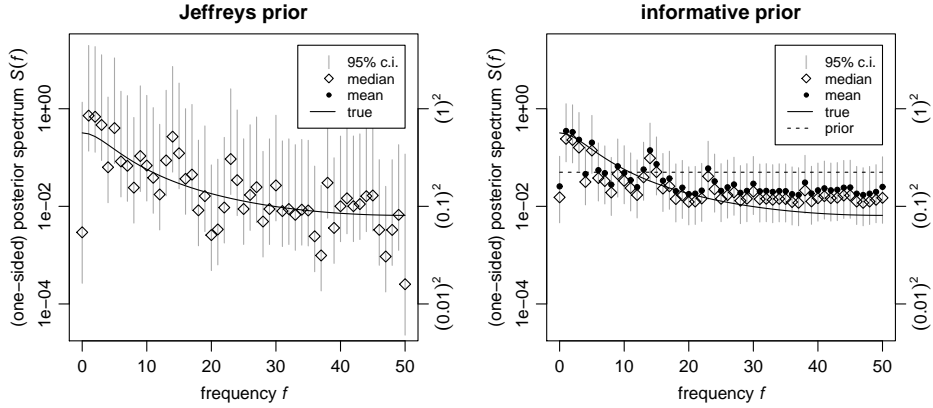}
    \caption{Posterior distributions of the 51 spectrum
             parameters~$\sigma_j^2$ based on the data shown in
             figure~\ref{fig:example01}. The left plot shows the posterior
             corresponding to the uninformative (and improper) Jeffreys
             prior. Posterior expectations do not exist in this
             case. The right plot corresponds to assuming a~priori white
             noise with $\nu_j=3$ degrees of freedom for each frequency
             bin; the dashed line marks the prior expectation value.}
    \label{fig:example02}
  \end{center}
\end{figure}
Figure~\ref{fig:example02} illustrates the resulting posterior
distribution for all 51 noise parameters~(\ref{eqn:sigmaposterior2})
in comparison to the case of using the uninformative (and improper)
Jeffreys prior.  The Jeffreys prior (with $\nu_j=0$ degrees of freedom
for each frequency bin) does not depend on the scale
parameters~$s_j^2$, and the resulting posterior with $\leq 2$ degrees
of freedom at each frequency does not have finite expectation values.
Note also that the posterior distributions corresponding to the first
and last frequency bin (zero and Nyquist frequencies, $\sigma_0^2$ and
$\sigma_{50}^2$) are wider than the others in both cases, as they have
one less degree of freedom.

\begin{figure}
  \begin{center}
    \includegraphics[width=0.95\columnwidth]{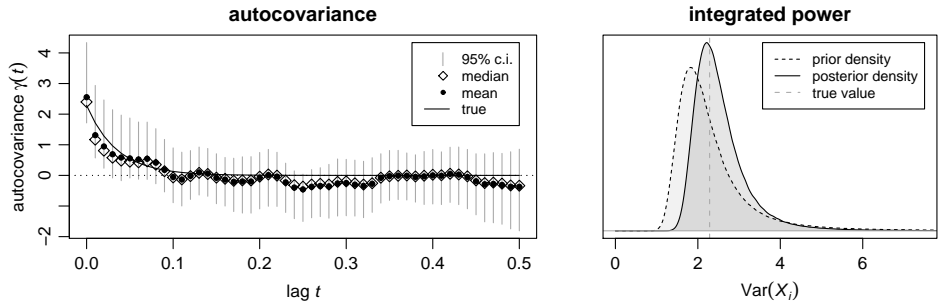}
    \caption{Posterior distributions of autocovariance and variance; the
             distributions shown here were derived via Monte Carlo
             integration.}
    \label{fig:example03}
  \end{center}
\end{figure}
Figure~\ref{fig:example03} shows the posterior distributions of
autocovariance and variance, which are functions of the individual
spectrum parameters~$\sigma^2_j$. The variance may either be
considered as the zero-lag autocovariance~$\acov(0)$
(see~(\ref{eqn:autocov})), or as the integrated power
$\intspec_{[0,f_{N/2}]}$ across the whole frequency range
(see~(\ref{eqn:intspec})).

\subsection{A signal with additive noise}\label{sec:example2}
In the following example we will consider a time series where the
primary interest is in a \textsl{signal} component, while the additive
\textsl{noise} component will be modelled using the approach
introduced above.  As in the previous section, we will again consider
$N=100$ data points $y(t_1),\ldots,y(t_{100})$ that are modelled as
\begin{equation}\label{eqn:MCMCmodel}
  y(t_i) \;=\; g_{f,\dot{f},a,\phi}(t_i) \,+\, n_{\vec{\sigma}}(t_i),
\end{equation}
where $n_{\vec{\sigma}}(t_i)$ is non-white noise of unknown spectrum,
and $g_{f,\dot{f},a,\phi}(t)$ is a ``chirping'' signal waveform of
increasing frequency:
\begin{equation}
  g_{f,\dot{f},a,\phi}(t) \;=\; a\, \sin(2\pi (f+\dot{f}t)t + \phi)
\end{equation}
where $f$ and $\dot{f}$ are the \textsl{frequency} and
\textsl{frequency derivative}, $a$~is the \textsl{amplitude}, and
$\phi$ is the \textsl{phase}.  The noise again is generated the same
way as in the previous example, by simulating an autoregressive
process.
The signal's ``size'' relative to the noise is given by the
signal-to-noise ratio (SNR: $\varrho =
\sqrt{4\sum_j\frac{\frac{\Delta_t}{N}|\tilde{g}(f_j)|^2}{\powspecC_1(f_j)}}$),
which here is at~15.

We define the signal parameters' prior as uniform for phase, frequency
and amplitude ($\phi\in[0,2\pi]$, $f\in[1,50]$ and $a\in[0,10]$) and
normal for the frequency derivative $\dot{f}$ (zero mean and standard
deviation~5).  The prior distribution for the noise parameters
$\sigma^2_0,\ldots,\sigma^2_{50}$ is set exactly as in the previous
section~\ref{sec:example2}.  For comparison, we analyze the data two
more times, once assuming the true noise spectrum to be known, and
once assuming the noise to be white, but with an unknown variance.  In
the latter case, we again assume a (conjugate) $\invchisq$ prior with
an expectation of 2.5 and 3 degrees of freedom for the variance
parameter.

The posterior distributions of signal and noise parameters may now be
derived via Monte Carlo integration; we implemented a Metropolis
sampler to simulate draws from the joint posterior probability
distribution of all parameters \cite{BDA}.  For the model including
the coloured noise spectrum as unknown, the signal parameters ($f$,
$\dot{f}$, $a$, $\phi$) may be sampled based on the marginal
likelihood expression~(\ref{eqn:margLikelihood6}), while samples of
the 51~noise parameters ($\sigma^2_0,\ldots,\sigma^2_{50}$) may then
be sampled in an additional step via the conditional distribution of
noise parameters for given signal parameters and the corresponding
implied vector of noise residuals~(\ref{eqn:sigmaposterior2}).
Similarly, for the known spectrum model, sampling of the four signal
parameters may be based on the likelihood
expression~(\ref{eqn:condLikelihood1}), while sampling for the white
noise model may be done using a Gibbs sampler alternately sampling
from the conditional distributions of four signal parameters and the
noise parameter~\cite{BDA}.
\begin{figure}
  \begin{center}
    \includegraphics[width=0.95\columnwidth]{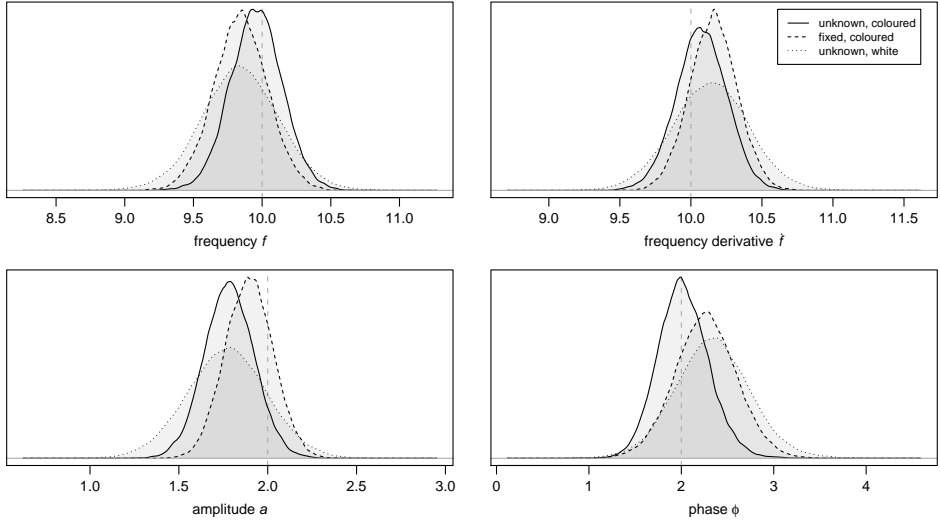}
    \caption{Marginal posterior densities of the four individual
             \textsl{signal} parameters.  The three densities in each
             panel result from three different models applied for the
             noise term. The ``\textsl{unknown, coloured}'' case
             corresponds to the model introduced above, the
             ``\textsl{fixed, coloured}'' case assumes the true noise
             spectrum to be a~priori known, and in the ``\textsl{unknown
             white}'' case the noise is modelled as white with unknown
             variance.  The vertical dashed lines indicate the true
             parameter values.}
    \label{fig:example04}
  \end{center}
\end{figure}

Figure~\ref{fig:example04} shows the resulting marginal posterior
distributions of the four signal parameters in comparison.  
\begin{figure}
  \begin{center}
    \includegraphics[width=0.95\columnwidth]{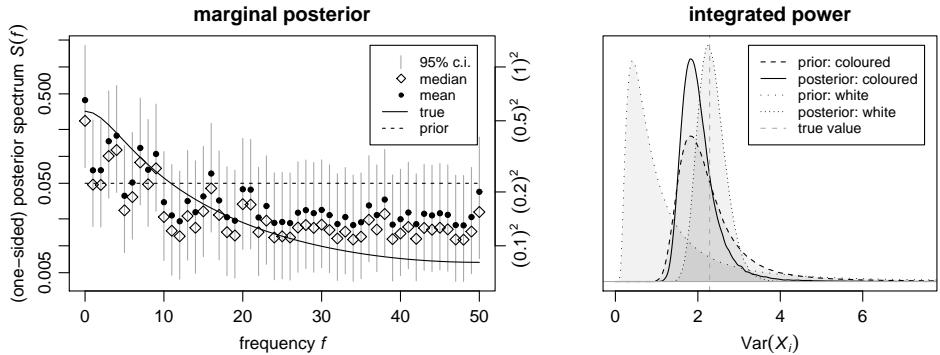}
    \caption{Marginal posterior distributions of the noise parameters.
             The left panel shows the posterior distribution for the 51
             individual parameters of the coloured noise model. The
             right panel shows the prior and posterior distributions of
             the integrated power in comparison with the corresponding
             variance parameter in the white noise model.}
    \label{fig:example05}
  \end{center}
\end{figure}
One can see that application of the more flexible coloured noise model
for the error term yields a more precise posterior distribution than
the white noise model, and in fact the posterior is more similar to
the case when the true noise spectrum is plugged into the model.
Looking at the (marginal) posterior distributions of the noise
parameters in figure~\ref{fig:example05}, one can see that although
the noise spectrum is only recovered with great uncertainty, this does
not seem to harm the estimation of the signal parameters of actual
interest.
The overall variance is better recovered by the single-parameter white
noise model, whereas the adaptability of the coloured noise model to
the predominantly low-frequency noise (which \textsl{is} reflected in
the posterior) seems to be the greater advantage.  While the relative
accuracy of the different methods is subject to a multitude of
circumstances, one can see that depite its considerably greater
complexity, the coloured noise model seems to perform competitively
here.

\section{Discussion}\label{sec:discussion}
This work originated out of the \textsl{Mock LISA Data Challenges
(MLDC)} \cite{MLDCoverview}, a gravitational wave parameter
estimation effort in the context of the planned \textsl{Laser
Interferometer Space Antenna (LISA)}\@. Here parameters of a signal were
to be inferred, where the signal was buried in noise known to be
non-white and interspersed with a host of individual emission lines
\cite{BarackCutler2004}.  So the problem was to model a non-white,
non-continuous spectrum where the shape of the spectrum was only
vaguely known in advance \cite{Roever2007-thesis,RoeverEtAl2007c}.
Also, the spectrum itself was not of primary interest, but rather a
nuisance parameter that still needed to be accounted for along with
the actual signal.  Application of the method described here solved
the problem and allowed the implementation 
of a Markov chain Monte Carlo (MCMC) algorithm
based on a straightforward generalization of the commonly utilized
Gaussian noise assumption, where modelling the spectrum complicated
the analysis only slightly.

Several approaches to the simultaneous estimation of signal and noise
parameters have been proposed before.  An obvious way to model
non-white noise would be to assume a particular time series
formulation that allows for some flexibility in the resulting
spectrum, for example an autoregressive (AR) model.  Representations
of this kind have been applied in various signal processing contexts,
e.g.\ for detecting sinusoidal signals and estimating their parameters
when these are buried in coloured noise
\cite{ChatterjeeEtAl1987,ChoDjuric1995}, or in the context of musical
pitch estimation, where the signals to be modelled again are
sinusoidal, but include higher harmonics and time-varying amplitudes
\cite{GodsillDavy2002}.  If one is only interested in a narrow
frequency band of the data, it may also be appropriate to model the
noise spectrum as constant across the concerned range
\cite{DupuisWoan2005}.  Regarding data treated in their Fourier
domain representation, there is a rich literature concerned with the
\textsl{measurement} of spectral densities (see e.g.\
\cite{Brillinger} and references therein).  However, there the aim is
usually to produce consistent and smooth estimates of the spectrum,
and the approaches applied include e.g.\ averaging
\cite{Welch1967,Harris1978}, smoothing via splines \cite{Wahba1980}
or Bayesian model fitting
\cite{GangopadhyayMallikDenison1999,ChoudhouriGhosalRoy2004b}.

The approach to modelling the noise spectrum introduced here is
different in that the spectrum \textsl{per~se} is not of interest, or
only of interest as far as it enters into likelihood computations.
Smoothness or interpolation therefore are not primarily aimed for.
What in fact is of concern is properly accounting for an a~priori
uncertainty in the (discretized, convolved) spectrum, in the frequency
resolution that is given by the numerically Fourier transformed data,
which then consequently entails the necessity for updating our
knowledge about the spectrum as data is being processed.  The
resulting approach generalizes the model underlying the commonly used
\textsl{Whittle likelihood}
\cite{Whittle1957,Finn1992,ChoudhouriGhosalRoy2004a}, and is hence
essentially based on the ``plain'' periodogram of the noise time
series, which for other purposes is commonly dismissed due to its
unfavourable large-sample convergence behaviour.  The model is very
flexible as it is built upon a ``binned'' spectrum estimate without
introducing any extra assumptions on the shape of the underlying noise
spectral density.  Its generality and simplicity make it useful for
modelling residual noise at very little computational cost.  
In the way it is defined,
the model is very general; it is a generalization of
the model that constitutes the basis for \textsl{matched filtering}
(e.g.\ \cite{Turin1960,WainsteinZubakov}) and that is commonly
applied in signal processing problems (e.g.\ \cite{Finn1992}), which
then in turn constitutes the special case of an a~priori known
spectrum.  The additional feature of the approach introduced here is
that it allows to specify corresponding uncertainties in addition to
the (prior) scale of the noise spectrum.  Marginalization over the
uncertainty in the noise spectrum yields a Student-$t$ model for the
Fourier-domain data as a natural generalization of the common normal
model.  Due to the straightforward interpretability of the model and
its computationally convenient form, we expect it to be particularly
useful for modelling an unknown noise spectrum that constitutes a
nuisance parameter rather than being of interest in itself.  In that
sense, the approach is not primarily aimed at gaining information
about an unknown spectrum, but rather at properly accounting for
uncertainty, and avoiding bias from supposedly precise a~priori
knowledge of the spectrum.

Alternatively, the Student-$t$ model may also be viewed as a
generalized, robust model accommodating for heavier-tailed,
non-Gaussian, or non-stationary noise.  Related approaches are
commonly used in many other applications in the context of
\textsl{robust statistics}
\cite{LangeEtAl1989,Geweke1993,Divgi1990,McDonaldNewey1988,Hampel,Huber}.
In fact, the use of similar methods for accommodating noise outliers
have already been advocated in the context of gravitational wave
signal processing \cite{Creighton1999,AllenEtAl2002}.  Along similar
lines, Clark et al.\ \cite{ClarkHengPitkinWoan2007} implemented a
(sine-Gaussian) noise-glitch component into the noise model.  Principe
and Pinto \cite{PrincipePinto2009} suggested the use of a dictionary
of glitch ``atoms'' in order to account for burst-like noise events.
Veitch and Vecchio \cite{VeitchVecchio2010} implemented a coherence
test based on a model selection procedure that is able to discriminate
noise glitches (that appear independently in the data) from actual
astrophysical events (that appear coherently in several data streams).
Similarly, Littenberg and Cornish \cite{LittenbergCornish2010} extend
their model to allow for excess noise that is isolated in time and
frequency via a wavelet approach.  Middleton \cite{Middleton1951abc}
treated the general case of Gaussian noise with superimposed
Poisson-distributed impulse noise bursts.  We are planning to
investigate the performance and sensitivity of a Student-$t$ model for
robust detection and parameter estimation in real interferometer
noise.

We expect the approach to be especially useful as a model component
properly accounting for non-white residual noise, at little
computational cost and without introducing overly restrictive
assumptions about the noise.  The interpretability of the Student-$t$
model in the context of robust modeling and M-estimation also makes it
straightforwardly useful for robust inference.  In that way, it may
particularly be useful in signal processing contexts
\cite{McDonoughWhalen,WainsteinZubakov}, but it should be applicable
in any case where a model for (residual) noise is
needed~\cite{JenkinsWatts,Bloomfield}.  The fact that the model
represents noise properties in terms of its power spectral density and
is specified through physically meaningful parameters may make it
particularly appealing in physical or engineering applications, where
modelling is commonly based on Fourier-domain descriptions, and a
time-domain formulation might be hard to incorporate or motivate.
The advantage from application of the more flexible
Student-$t$ model will very much depend on the particular inference
problem at hand. The common normal model will obviously be optimal
when its assumptions are met, and the degree to which one will
outperform the other will depend on the particular departure(s) from
Gaussianity, or on the imperfect prior knowledge of the spectrum, even
if the data are perfectly Gaussian.  While the range of possible
deviations from the standard Gaussian model is infinite, some insight
into the behaviour of a Student-$t$ model may be gained from the
\textsl{robust statistics} literature; the properties of such an
approach for location estimation has been investigated via Monte
Carlo studies \cite{PrincetonRobustnessStudy,Divgi1990} as well as
theoretical considerations of key figures like \textsl{relative
efficiency} or \textsl{breakdown point} \cite{Divgi1990,Hampel,Huber}.
A case study in the regression context may be found in \cite{Geweke1993}.

The above basic model may in future be extended by introducing
smoothness constraints on the spectrum. This might for example be
approached by considering correlations between neighbouring spectral
bins, or by assuming a piecewise constant spectrum, similar to what
was done in \cite{DupuisWoan2005}; the latter approach would in fact
be a compromise between two models used in the example discussed above
(Sec.~\ref{sec:example2}), 
namely a flat spectrum and individually
modelled frequency bins.  Another interesting extension would be the
incorporation of cross-spectra
\cite{JenkinsWatts,NofrariasEtAl2010}.  Some of the methods
described here have been coded as an R~software extension package and
are available for free download \cite{bspec}.

\ack
This work was supported by 
the Max-Planck-Society, 
The Royal Society of New Zealand Marsden Fund grant \mbox{UOA-204}, 
and National Science Foundation grants \mbox{PHY-0553422} and \mbox{PHY-0854790}.

\appendix
\section*{Appendix}
\setcounter{section}{1}
\subsection{Discrete Fourier transform (DFT)}\label{sec:DFTdefinition}
The Fourier transform convention used in this paper is specified
below; it is defined for a real-valued function~$h$ of time~$t$,
sampled at $N$ discrete time points, at a sampling rate of
$\frac{1}{\Delta_t}$, and it maps from
\begin{equation} \label{eqn:DFTtimedomain}
  \{h(t)\in\realline:\; t=0,\Delta_t,2\Delta_t,\ldots,(N-1)\Delta_t\}
\end{equation}
to a function of frequency~$f$
\begin{equation} \textstyle
  \{\tilde{h}(f)\in\complexnumb:\; f=0,\Delta_f,2\Delta_f,\ldots,(N-1)\Delta_f\},
\end{equation}
where $\Delta_f = \frac{1}{N\Delta_t}$ and
\begin{equation}\label{eqn:DftDefinition}
  \tilde{h}(f) \;=\; \sum_{j=0}^{N-1} h(j\Delta_t) \exp(-2 \pi \imag j \Delta_t f).
\end{equation}
The inverse DFT then is given by
\begin{equation}\label{eqn:InvDftDefinition}
  h(t) \;=\; \frac{1}{N} \sum_{j=0}^{N-1} \tilde{h}(j\Delta_f) \exp(2 \pi \imag j \Delta_f t)
\end{equation}
\cite{Gregory}.

\subsection{Relationship between DFT and time series model}\label{sec:DFTrelation}
Let
\begin{equation}
 \alpha_j\;=\;\re\bigl(\tilde{h}(f_j)\bigr) \quad \mbox{and} \quad
 \beta_j\;=\;\im\bigl(\tilde{h}(f_j)\bigr),
\end{equation}
i.e.: $\tilde{h}(f_j)=\alpha_j+\beta_j\imag$.
For simplicity, in the following $N$ is assumed to be even; for
uneven~$N$ the derivation is similar.
The inverse~DFT was defined as (\ref{eqn:InvDftDefinition}):
\begin{eqnarray}
 h(t) &=& \frac{1}{N} \sum_{j=0}^{N-1}\tilde{h}(f_j)\exp(2\pi\imag f_j t)\\[1ex]
     &=& \frac{1}{N} \sum_{j=1}^{\frac{N}{2}-1}\Bigl[
          \bigl(2\alpha_j\cos(2\pi f_j t) + 2(-\beta_j)\sin(2\pi f_j t)\bigr) 
          \Bigr]\nonumber\\
      &&  +\, \textstyle\frac{1}{N}\alpha_0 
          \,+\, \textstyle\frac{1}{N}\alpha_{N/2}\cos(2\pi\imag f_{N/2} t)
\end{eqnarray}
where $t\in\{0,\Delta_t,2\Delta_t,\ldots,(N-1)\Delta_t\}$, and
$f_j=j\Delta_f=\frac{j}{N \Delta_t}$ are the Fourier frequencies.  So,
comparing to (\ref{eqn:SinusoidalNoise}), one can see that the
realizations of $a_0,\ldots,a_{\lfloor N/2\rfloor}$ and
$b_0,\ldots,b_{\lfloor N/2\rfloor}$ are derived from a given time
series by Fourier-transforming and then setting
\begin{equation}\label{eqn:DFTrelation1}
  a_j \;=\; \textstyle  \indi{j} \,\sqrt{\frac{\Delta_t}{N}}\,  \alpha_j \quad \mbox{and} \quad
  b_j \;=\; \textstyle -\indi{j} \,\sqrt{\frac{\Delta_t}{N}}\,  \beta_j
\end{equation}
for $j=0,\ldots,\lfloor N/2\rfloor$, which especially implies that
\begin{equation}\label{eqn:DFTrelation2}
  a_j^2 + b_j^2 \;=\; \textstyle \indi{j}^2 \, \frac{\Delta_t}{N} \, (\alpha_j^2 + \beta_j^2)
  \;=\; \textstyle \indi{j}^2 \, \frac{\Delta_t}{N} \, \bigl|\tilde{h}(f_j)\bigr|^2.
\end{equation}

\section*{References}
  \bibliographystyle{unsrt}
  \bibliography{../../literature/literature}

\end{document}